\documentclass{llncs}
\usepackage{psfig,url,makeidx,longtable}
\usepackage{epsf,epsfig,ifthen,array,latexsym,url}
\usepackage{amsmath,amstext,amsfonts,longtable,amssymb}
\usepackage{psfrag}

\psfrag{s0}{{\scriptsize$s_0$}}
\psfrag{s1}{{\scriptsize$s_1$}}
\psfrag{s2}{{\scriptsize$s_2$}}
\psfrag{s3}{{\scriptsize$s_3$}}
\psfrag{s4}{{\scriptsize$s_4$}}
\psfrag{s5}{{\scriptsize$s_5$}}
\psfrag{s6}{{\scriptsize$s_6$}}
\psfrag{s7}{{\scriptsize$s_7$}}
\psfrag{s8}{{\scriptsize$s_8$}}
\psfrag{oB}{{$\o B$}}

\newtheorem{prop}{Property}

\newlength{\tmplength}
\newenvironment{lawhint}%
  {\setlength{\tmplength}{0.5\linewidth}\addtolength{\tmplength}{1cm}%
    \begin{longtable}[l]{>{$}p{\tmplength}<{$}l}}%
  {
   \end{longtable}}

  {\setlength{\tmplength}{0.5\linewidth}\addtolength{\tmplength}{1cm}%
   \begin{flushleft}\begin{tabular}[l]{>{$}p{\tmplength}<{$}l}}%
  {\end{tabular}\end{flushleft}}

  {\textbf{Proof:}\newline\begin{sffamily}\small}%
  {\end{sffamily}\hfill$\blacksquare$

  }
  {\setlength{\tmplength}{0.5\linewidth}\addtolength{\tmplength}{0.3cm}%
   \begin{flushleft}\begin{tabular}[l]{>{$}p{0.7cm}<{$}@{}>{$}p{\tmplength}<{$}l}}%
  {\end{tabular}\end{flushleft}}

  {\begin{longtable}[l]{>{$}p{0.7cm}<{$}@{}>{$}p{12cm}<{$}l}}%
  {\end{longtable}}
\newenvironment{widestproof}%
  {\begin{longtable}[l]{>{$}p{0.7cm}<{$}@{}>{$}l<{$}}}%
  {\end{longtable}}
  {\setlength{\tmplength}{0.5\linewidth}\addtolength{\tmplength}{0.3cm}%
   \begin{longtable}[l]{>{$}p{0.7cm}<{$}@{}>{$}p{\tmplength}<{$}l}}%
  {\end{longtable}}

\newcommand{\lhint}[1]{\\ &\rule{0cm}{3ex}\text{#1} \\ }
        
\newcommand{\llawscr}{\end{math}\newline\begin{math}}
\newcommand{\llawsname}[1]{\end{math}#1\newline\begin{math}}
{\let\tmpa=\\\let\\=\llawscr\let\tmpb=\name\let\name=\llawsname%
 \setlength{\columnseprule}{0pt} %
  \ifthenelse{\equal{#1}{}}{}{#1\vspace{-0.3cm}} %
 \begin{multicols}{2}\begin{math}} %
{\end{math}\let\\=\tmpa\let\name=\tmpb\end{multicols}}

\newcommand{\nm}{\mathit}

\newcommand{\eq}{=}
\newsavebox{\difsign}
\savebox{\difsign}{${|}\!\!\!{=}$}

\newcommand{\ltop}{\top}
\newcommand{\lbot}{\bot}
\newcommand{\limp}{\Rightarrow}
\newcommand{\lpmi}{\Leftarrow}
\renewcommand{\leq}{\eq}

\newcommand{\lif}[3]{\mathord{#2 \lhd #1 \rhd #3}}
\newcommand{\Lif}[3]{\mathord{%
  \mathbf{if}\;#1\;%
  \mathbf{then}\;#2\;%
  \mathbf{else}\;#3}%
}
\newcommand{\largeboldbinop}[1]{\mathbin{\mbox{\large$\boldsymbol{#1}$}}}

\newcommand{\Lpmi}{\largeboldbinop{\lpmi}} 
\newcommand{\Leq}{\largeboldbinop{\leq}}

\newcommand{\nat}{\mathbb{N}}

\newcommand{\quant}[3]{\ifthenelse{\equal{#3}{}}{\mathord{#1} #2 \cdot}{\mathord{#1} #2\in #3 \cdot}}
\newcommand{\A}{\quant{\forall}}   

\newcommand{\stseq}{\nm{stseq}}
\newcommand{\llbrack}{\mathopen{\lbrack\!\lbrack}}
\newcommand{\rrbrack}{\mathclose{\rbrack\!\rbrack}}
\newcommand{\interp}[2]{#2 \llbrack #1 \rrbrack}

\renewcommand{\a}{\mathord{\Box}}               
\newcommand{\e}{\mathord{\Diamond}}             
\newcommand{\U}{\mathbin{\mathcal{U}}}          
\newcommand{\W}{\mathbin{\mathcal{W}}}          
\renewcommand{\P}{\mathbin{\mathcal{P}}}        
\renewcommand{\o}{\mathord{\circ}}              
\newcommand{\eu}{\mathord{\uparrow}}            
\newcommand{\ed}{\mathord{\downarrow}}          
\newcommand{\ea}{\mathord{\updownarrow}}        
\newcommand{\cusleft}{\mathopen{\ll}}
\newcommand{\cusright}{\mathclose{\gg}}
\newcommand{\cus}[1]{\cusleft #1 \cusright}

\begin{document}
\frontmatter
\title{Events in Property Patterns}

\author{Marsha Chechik \and Dimitrie O. P\u{a}un}
\institute{Department of Computer Science,
University of Toronto,\\Toronto, ON M5S 3G4, Canada.\\
Email: \email{\{chechik,dimi\}@cs.toronto.edu}}
\date{\today}

\maketitle
\begin{abstract}
A pattern-based approach to the presentation, codification and reuse
of property specifications for finite-state verification was proposed
by Dwyer and his colleagues in~\cite{dwyer99,dwyer98a}.  The patterns
enable non-experts to read and write formal specifications for
realistic systems and facilitate easy conversion of specifications
between formalisms, such as LTL, CTL, QRE.  In this paper we extend
the pattern system with {\em events} --- changes of values of
variables in the context of LTL.

\end{abstract}
\section{Introduction}
\label{intro}
Temporal logics (TL) (e.g.,~\cite{bernstein81}, \cite{hailpern83},
\cite{lamport83}, \cite{ostroff90}, \cite{manna92}) have received a
lot of attention in the research community.  Not only are
they useful for specifying properties of systems, recent advances
in model-checking allow effective automatic checking of models of systems
against such properties, e.g. using tools like SPIN~\cite{holzmann97}
and SMV~\cite{mcmillan93}.

One important obstacle to using temporal logic is the ability to
express complex properties correctly.  To remedy this problem, Dwyer
and his colleagues have proposed a pattern-based approach to the
presentation, codification and reuse of property specifications.  The
system allows patterns like ``$P$ is absent between $Q$ and $S$'' or
``$S$ precedes $P$ between $Q$ and $R$'' to
be easily expressed in and translated between 
linear-time temporal logic (LTL)~\cite{manna90}, computational tree
logic (CTL)~\cite{clarke86}, quantified regular expressions
(QRE)~\cite{oleander90}, and other state-based and event-based
formalisms.  Dwyer et. al. also performed a large-scale study in which
specifications containing over 500 temporal properties were collected
and analyzed. They noticed that over 90\% of these could be classified
under one of the proposed patterns~\cite{dwyer99}.

In earlier work~\cite{paun98}, we used the Promela/SPIN framework
to model the Production Cell system.  We attempted to use the
pattern-base approach to help us formalize properties of this system
in LTL.  However, we found that the approach could not be applied
directly, because our properties used {\em events} --- changes of
values of variables, e.g., ``magnet should become deactivated'', which
we wanted to formalize as ``magnet is active now and will be inactive
in the next state''.  We called such events {\em edges}.

LTL is a temporal logic comprised of propositional formulas and
temporal connectives $\a$ (``always''), $\e$ (``eventually''), $\o$ (``next''),
and $\U$ (``until''). The first three operators are unary, while the last
one is binary.  $\U$ is the {\em strong until}; that is, it requires
that $B$ actually happen sometime in the future. In this context, we
define edges as follows:
\[
\begin{array}{rcl>{$}c<{$}>{$}l<{$}}
\eu A &\Leq& \lnot A \land\o A  & --- & up or rising edge    \label{eq:eu} \\
\ed A &\Leq& A \land \o\lnot A  & --- & down or falling edge \label{eq:ed} \\
\ea A &\Leq& \eu A \lor \ed  A  & --- & any edge             \label{eq:ea}
\end{array}
\]

LTL formulas containing events may have problems caused by the use of
the ``next'' operator in the definition of edges.  Temporal formulas
that make use of ``next'' may not be {\em closed under stuttering},
i.e.  their interpretation may be modified by transitions that leave
the system in the same state (``stutter''). As we discuss later in the
paper, this is an essential property for effective use of temporal
formulas.

Model-checking allows relatively novice users to
verify correctness of their systems quickly and effectively.  However,
it is essential that these users are able to specify correctness
criteria in the appropriate temporal logic.  For example, effective
use of SPIN~\cite{holzmann97} depends critically on being able to
express such criteria in LTL.  Under the presence of events, it is
often quite complex (see ~\cite{paun99a} for a thorough discussion).
In this paper we extend the properties of Dwyer et. al. to include
events in LTL properties. The rest of the paper is organized as
follows: Section~\ref{patternsystem} overviews the pattern-based
system.  Section~\ref{extensions} presents our extension to the
pattern-based system and discusses the extension
process. Section~\ref{cus} contains an informal summary of our
treatment of closure under stuttering and presents a set of theorems
that allow syntactic checking of formulas for this property. In
addition, it shows how to use these theorems to prove that our
extensions of the pattern-based system are closed under stuttering.
Section~\ref{conclusion} concludes the paper.

\section{Overview of the Pattern-Based Approach}
\label{patternsystem}
\begin{figure*}[t]
\begin{center}
\psfig{file=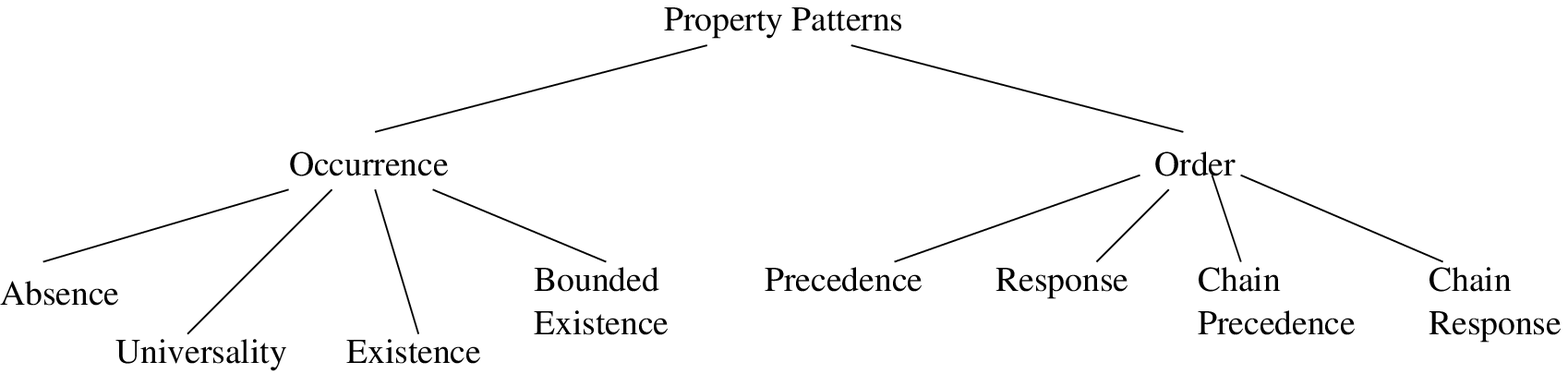,width=\linewidth}
\caption{A Pattern Hierarchy.}
\label{fig:hierarchy}
\end{center}
\end{figure*}

In this section we survey the pattern-based approach.  For more
information, please refer to~\cite{dwyer98a,dwyer99}.
The patterns are organized hierarchically based on their semantics, as
illustrated in Figure~\ref{fig:hierarchy}.  Some of the patterns are
described below:
\begin{tabbing}
{\bf Universality} \= blah \kill
{\bf Absence} \> A condition does not occur within a scope;\\
{\bf Existence} \>  A condition must occur within a scope;\\
{\bf Universality} \>  A condition occurs throughout a scope;\\
{\bf Response} \>  A condition must always be followed by another
  within a scope;\\
{\bf Precedence} \> A condition must always be preceded by
  another within a scope.
\end{tabbing}

\begin{figure}[htb]
\begin{center}
\psfig{file=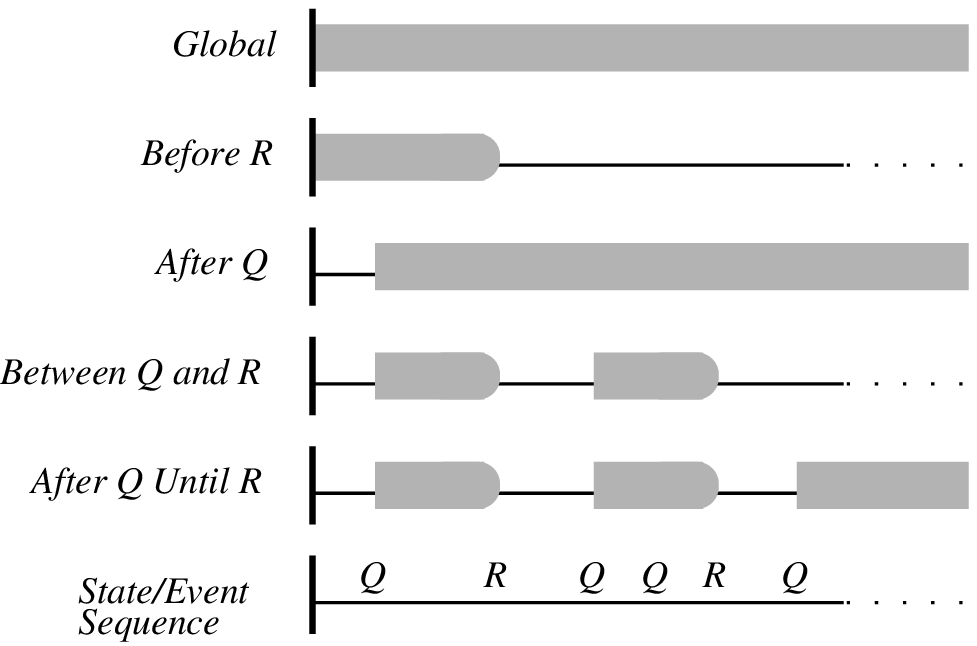,width=2.5in} 
\caption{Pattern Scopes.}
\label{fig:scopes}
\end{center}
\end{figure}

Each pattern is associated with several \emph{scopes} --- the regions
of interest over which the condition is evaluated.
There are five basic kinds of scopes:
\begin{tabbing}
{\bf D.} \=  {\bf Between $Q$ and $R$} \= \kill
{\bf A.} \> {\bf Global} \>     The entire state sequence;\\
{\bf B.} \> {\bf Before $R$} \>  The state sequence up to condition $R$;\\
{\bf C.} \> {\bf After $Q$} \>  The state sequence after condition $Q$;\\
{\bf D.} \> {\bf Between $Q$ and $R$} \> The part of the state sequence between
    condition $Q$ and\\
\>  condition $R$;\\
{\bf E.} \> {\bf After $Q$ Until $R$} \> Similar to the previous one, except
    that the designated\\ \> part of the state sequence continues even if the
    second condition does not\\ \> occur.
\end{tabbing}
These scopes are depicted in Figure~\ref{fig:scopes}. The scopes were
initially defined in~\cite{dwyer99} to be closed-left, open-right
intervals, although it is also possible to define other combinations,
such as open-left, closed-right intervals. 

For example, an LTL formulation of the property ``$S$ precedes $P$
between $Q$ and $R$'' (\textbf{Precedence} pattern with ``between $Q$
and $R$'' scope) is:
$$ \a((Q \land \e R) \limp (\lnot P \U (S \lor R))) $$
Even though the pattern system is
formalism-independent~\cite{dwyer98a}, in this paper we are
only concerned with the expression of properties in LTL.

S\section{Edges and the Pattern-Based System}
\label{extensions}

LTL is a state-based formalism, and thus the original pattern
system does not specify the expression of events in LTL.
In this section we show how to include reasoning about events
to the pattern system.  These events can be used for specifying
conditions as well as for defining the bounding scopes.

We start by introducing some notation that allows us a more compact
representation of properties.  We define the {\em weak} version of
``until'' as:
\[
  A \W B = \a A \lor (A \U B)
\]
That is, we no longer require $B$ to happen in the future; if it does
not, than $A$ should hold indefinitely.  Another useful operator is
``precedes'':
\[
  A \P B = \lnot(\lnot A \U B)
\]
That is, we want $A$ to hold \emph{before} $B$ does. Note that in this
case $B$ may never happen.  Also, we use $\lif{x}{y}{z}$ to indicate
$\Lif{x}{y}{z}$, or $(x \land y) \lor (\lnot x \land z)$.  Finally, we
write $\top$ and $\bot$ to indicate boolean values {\em true} and {\em
false}, respectively.

Since our extension involves edges, we give a few relevant
properties below:
\begin{eqnarray}
  \a\eu A &=& \bot  \label{eq:Aeu} \\
  \a\ed A &=& \bot  \label{eq:Aed} \\
  \eu A \U B &=& B \lor (\eu A \land \o B) \label{eq:Ueu}
\end{eqnarray}
Properties (\ref{eq:Aeu}) and (\ref{eq:Aed}) indicate that edges
of the same type cannot occur in every state of the system, whereas
property (\ref{eq:Ueu}) allows us to replace an
``until'' with a propositional expression.  

We have explored the concept of edges in~\cite{paun99a,paun99b}, and
list some of edge properties in the Appendix.

\subsection{Extending the Pattern System}

Introducing edges into the patterns generates an explosion in the
number of formulas: conditions can be state-based or edge-based,
inclusive or exclusive, and the interval ends can be either opened or
closed. Our extension does not include all the possibilities, but
rather a significant and representative set of them, as discussed
below.


We were able to extend five of the nine patterns: \textbf{Absence}
(Figure~\ref{fig:patAb}), \textbf{Existence} (Figure~\ref{fig:patE}),
\textbf{Universality} (Figure~\ref{fig:patA}), \textbf{Response}
(Figure~\ref{fig:patR}), \textbf{Precedence} (Figure~\ref{fig:patP}).
For each of the five scopes, we list four formulas corresponding to
the four combinations of state-based and edge-based conditions and
interval bounds we have considered:
\begin{center}
  \begin{tabular}{cll}
    0. & $P$, $S$ --- states,   & $Q$, $R$ --- states;       \\
    1. & $P$, $S$ --- states,   & $Q$, $R$ --- up edges;     \\
    2. & $P$, $S$ --- up edges, & $Q$, $R$ --- states;       \\
    3. & $P$, $S$ --- up edges, & $Q$, $R$ --- up edges.
  \end{tabular}  
\end{center}
Combination 0 corresponds to the original formulation of
\cite{dwyer98a}, where all of $P$, $S$, $Q$ and $R$ are state-based.
The remaining three combinations are our extensions to the pattern
system.  We assume that multiple events can happen simultaneously, but
only consider closed-left, open-right intervals, as in the original
system.  Also, we consider events $P$ and $S$ to be exclusive in
the \textbf{Precedence} pattern and inclusive in the \textbf{Response}
pattern\footnote{Two events are considered exclusive if they are not
allowed to happen at the same time, and inclusive otherwise.}. We note,
however, that it is perfectly possible to have formulas for all other
combinations of interval bounds.  Down edges can be substituted for up
edges without changing the formulas.  We have modified several of the
0-formulas (i.e.  state-based conditions and intervals) from their
original formulations of \cite{dwyer98a} to remove assumptions of
interleaving and make them consistent with the closed-left, open-right
intervals. Note that in the case of the \textbf{Universality} pattern,
we do not list formulas for edge-based events as edges cannot be
universally present (by (\ref{eq:Aeu}) and (\ref{eq:Aed})).

\begin{figure*}[t]
  \begin{center}
    \fbox{
      \begin{minipage}{\linewidth}
        \setlength{\arraycolsep}{2pt}
        \(
        \begin{array}{rl}
          \multicolumn{2}{c}{$A. Globally$} \\
          0.&\a\lnot P \\
          1.&\a\lnot P \\
          2.&\a\lnot\eu P \\
          3.&\a\lnot\eu P
        \end{array}
        \)
        \hfill
        \(
        \begin{array}{rl}
          \multicolumn{2}{c}{$B. Before \(R\)$} \\
          0.&\e R \limp (\lnot P \U R) \\
          1.&\e\eu R \limp (\eu R \P P) \\
          2.&\e R \limp (\lnot\eu P \U R) \\
          3.&\e\eu R \limp (\lnot\eu P \U \eu R)
        \end{array}
        \)
        \hfill
        \(
        \begin{array}{rl}
          \multicolumn{2}{c}{$C. After \(Q\)$} \\
          0.&\a(Q \limp \a\lnot P) \\
          1.&\a(\eu Q \limp\o \a\lnot P) \\
          2.&\a(Q \limp \a\lnot\eu P) \\
          3.&\a(\eu Q \limp \a\lnot\eu P)
        \end{array}
        \) 
        \vspace{3mm}\par\noindent
        \(
        \begin{array}{rl}
          \multicolumn{2}{c}{$D. Between \(Q\) and \(R\)$} \\
          0.&\a((Q \land \e R) \limp (\lnot P \U R)) \\
          1.&\a((\eu Q\land\e\eu R\land\lnot\eu R) \limp \o(\eu R \P P)) \\
          2.&\a((Q \land \e R) \limp (\lnot\eu P \U R)) \\
          3.&\a((\eu Q\land\e\eu R) \limp (\lnot\eu P\U\eu R))
        \end{array}
        \)
        \hfill
        \(
        \begin{array}{rl}
          \multicolumn{2}{c}{$E. After \(Q\) Until \(R\)$} \\
          0.&\a((Q \land \e P) \limp (\lnot P \U R)) \\
          1.&\a((\eu Q \land \lnot\eu R \land \o\e P) \limp \o(\eu R \P P)) \\
          2.&\a(Q \limp (\lnot\eu P \W R)) \\
          3.&\a(\eu Q \limp (\lnot\eu P \W \eu R))
        \end{array}
        \)
      \end{minipage}
      }
      \caption{Formulations of the \textbf{Absence} Pattern}
    \label{fig:patAb}
  \end{center}
\end{figure*}

\begin{figure*}[t]
  \begin{center}
    \fbox{
      \begin{minipage}{\linewidth}
        \setlength{\arraycolsep}{2pt}
        \(
        \begin{array}{rl}
          \multicolumn{2}{c}{$A. Globally$} \\
          0.&\e P \\
          1.&\e P \\
          2.&\e\eu P \\
          3.&\e\eu P
        \end{array}
        \)
        \hfill
        \(
        \begin{array}{rl}
          \multicolumn{2}{c}{$B. Before \(R\)$} \\
          0.&\e R \limp (P \P R) \\
          1.&\e\eu R \limp (\lnot\eu R \U P) \\
          2.&\e R \limp (\eu P \P R) \\
          3.&\e\eu R \limp (\eu P \P \eu R)
        \end{array}
        \)
        \hfill
        \(
        \begin{array}{rl}
          \multicolumn{2}{c}{$C. After \(Q\)$} \\
          0.&\e Q \limp \e(Q \land \e P) \\
          1.&\e\eu Q \limp \e(\eu Q \land \o\e P) \\
          2.&\e Q \limp \e(Q \land \e\eu P) \\
          3.&\e\eu Q \limp \e(\eu Q \land \e\eu P)
        \end{array}
        \) 
        \vspace{3mm}\par\noindent
        \(
        \begin{array}{rl}
          \multicolumn{2}{c}{$D. Between \(Q\) and \(R\)$} \\
          0.&\a((Q \land \e R) \limp ((P \P R) \land \lnot R)) \\
          1.&\a((\eu Q\land \e\eu R)\limp(\o(\lnot\eu R\U P)\land\lnot\eu R))\\
          2.&\a((Q \land\e R) \limp ((\eu P \P R) \land \lnot R)) \\
          3.&\a((\eu Q\land \e\eu R) \limp ((\eu P \P \eu R)\land\lnot\eu R))
        \end{array}
        \)
        \hfill
        \(
        \begin{array}{rl}
          \multicolumn{2}{c}{$E. After \(Q\) Until \(R\)$} \\
          0.&\a(Q \limp \lif{\e R}{(P \P R)\land \lnot R}{\e P}) \\
          1.&\a(\eu Q \limp \o(\lnot\eu R \U P) \land \lnot\eu R) \\
          2.&\a(Q\limp\lif{\e R}{(\eu P \P R)\land\lnot R}{\e\eu P}) \\
          3.&\a(\eu Q \limp  \lif{\e\eu R}
                      {(\eu P \P \eu R)\land \lnot\eu R}{\e\eu P})
        \end{array}
        \)
      \end{minipage}
      }
      \caption{Formulations of the \textbf{Existence} Pattern}
    \label{fig:patE}
  \end{center}
\end{figure*}

\begin{figure*}[t]
  \begin{center}
    \fbox{
      \begin{minipage}{\linewidth}
        \setlength{\arraycolsep}{2pt}
        \(
        \begin{array}{rl}
          \multicolumn{2}{c}{$A. Globally$} \\
          0.&\a P \\
          1.&\a P
        \end{array}
        \)
        \hfill
        \(
        \begin{array}{rl}
          \multicolumn{2}{c}{$B. Before \(R\)$} \\
          0.&\e R \limp (P \U R) \\
          1.&\e\eu R \limp (\eu R \P \lnot P)
        \end{array}
        \)
        \hfill
        \(
        \begin{array}{rl}
          \multicolumn{2}{c}{$C. After \(Q\)$} \\
          0.&\a(Q \limp \a P) \\
          1.&\a(\eu Q \limp\o \a P)
        \end{array}
        \) 
        \vspace{3mm}\par\noindent
        \(
        \begin{array}{rl}
          \multicolumn{2}{c}{$D. Between \(Q\) and \(R\)$} \\
          0.&\a((Q \land \e R) \limp (P \U R)) \\
          1.&\a((\eu Q\land\e\eu R\land\lnot\eu R)\limp\o(\eu R \P \lnot P))
        \end{array}
        \)
        \hfill
        \(
        \begin{array}{rl}
          \multicolumn{2}{c}{$E. After \(Q\) Until \(R\)$} \\
          0.&\a(Q \limp P \W R) \\
          1.&\a(\eu Q \limp \o \lif{\e\eu R}{(\eu R \P \lnot P)}{\a P})
        \end{array}
        \)
      \end{minipage}
      }
      \caption{Formulations of the \textbf{Universality} Pattern}
    \label{fig:patA}
  \end{center}
\end{figure*}

\begin{figure*}[t]
  \begin{center}
    \fbox{
      \begin{minipage}{\linewidth}
        \setlength{\arraycolsep}{2pt}
        \(
        \begin{array}{rl}
          \multicolumn{2}{c}{$A. Globally$} \\
          0.&\e P \limp S \P P \\
          1.&\e P \limp S \P P \\
          2.&\e\eu P \limp \eu S \P \eu P \\
          3.&\e\eu P \limp \eu S \P \eu P
        \end{array}
        \)
        \hfill
        \(
        \begin{array}{rl}
          \multicolumn{2}{c}{$E. After \(Q\) Until \(R\)$} \\
          0.&\a(Q \limp (\e P \limp 
                \lnot P \U ((S \land \lnot P) \lor R))) \\
          1.&\a(\eu Q \limp \o(\e P \limp 
                ((\lnot\eu R \U P) \limp (S \P P)))) \\
          2.&\a(Q \limp (\e P \limp 
                \lnot\eu P \U ((\eu S \land \lnot\eu P) \lor R))) \\
          3.&\a(\eu Q \limp \o(\e P \limp 
                ((\eu P \P \eu R) \limp (\eu S \P \eu P))))
        \end{array}
        \)
        \vspace{3mm}\par\noindent
        \(
        \begin{array}{rl}
          \multicolumn{2}{c}{$B. Before \(R\)$} \\
          0.&\e R \limp (\lnot P \U ((S \land \lnot P) \lor R)) \\
          1.&\e\eu R \limp ((\lnot\eu R \U P) \limp (S \P P)) \\
          2.&\e R \limp \lnot\eu P \U ((\eu S \land \lnot\eu P) \lor R) \\
          3.&\e\eu R \limp((\eu P \P \eu R) \limp (\eu S \P \eu P))
        \end{array}
        \)
        \hfill
        \(
        \begin{array}{rl}
          \multicolumn{2}{c}{$C. After \(Q\)$} \\
          0.&\e    Q \limp \e(    Q \land   (\e P    \limp (S \P P))) \\
          1.&\e\eu Q \limp \e(\eu Q \land \o(\e P    \limp (S \P P))) \\
          2.&\e    Q \limp \e(    Q \land   (\e\eu P \limp (\eu S \P \eu P)))\\
          3.&\e\eu Q \limp \e(\eu Q \land   (\e\eu P \limp (\eu S \P \eu P)))
        \end{array}
        \)
        \vspace{3mm}\par\noindent
        \hfill
        \(
        \begin{array}{rl}
          \multicolumn{2}{c}{$D. Between \(Q\) and \(R\)$} \\
          0.&\a((Q \land \e R) \limp 
                (\lnot P \U ((S \land \lnot P) \lor R))) \\
          1.&\a((\eu Q \land \lnot\eu R \o\e\eu R) \limp
                \o((\lnot\eu R \U P) \limp (S \P P))) \\
          2.&\a((Q \land \e R) \limp 
                (\lnot\eu P \U ((\eu S \land \lnot\eu P) \lor R))) \\
          3.&\a((\eu Q \land \lnot\eu R \o\e\eu R) \limp
                \o((\eu P \P \eu R) \limp (\eu S \P \eu P)))
        \end{array}
        \)
        \hfill
        \(\begin{array}{l}\rule{0cm}{1mm}\end{array}\) 
      \end{minipage}
      }
      \caption{Formulations of the \textbf{Precedence} Pattern}
    \label{fig:patP}
  \end{center}
\end{figure*}

\begin{figure*}[t]
  \begin{center}
    \fbox{
      \begin{minipage}{\linewidth}
        \setlength{\arraycolsep}{2pt}
        \(
        \begin{array}{rl}
          \multicolumn{2}{c}{$A. Globally$} \\
          0.&\a(P \limp \e S) \\
          1.&\a(P \limp \e S) \\
          2.&\a(\eu P \limp \e\eu S) \\
          3.&\a(\eu P \limp \e\eu S)
        \end{array}
        \)
        \hfill
        \(
        \begin{array}{rl}
          \multicolumn{2}{c}{$B. Before \(R\)$} \\
          0.&\e    R \limp (P     \limp (\lnot    R \U     S)) \U     R \\
          1.&\e\eu R\limp((P\limp(\lnot\eu R \U S))\land\lnot\eu R) \U
                          (\eu R \land (P \limp Q)) \\
          2.&\e    R \limp (\eu P \limp (\lnot    R \U \eu S)) \U     R \\
          3.&\e\eu R \limp (\eu P \limp (\lnot\eu R \U \eu S)) \U \eu R
        \end{array}
        \)
        \hfill
        \vspace{3mm}\par\noindent
        \(
        \begin{array}{rl}
          \multicolumn{2}{c}{$C. After \(Q\)$} \\
          0.&\a(Q     \limp   \a(    P \limp \e    S)) \\
          1.&\a(\eu Q \limp \o\a(    P \limp \e    S)) \\
          2.&\a(Q     \limp   \a(\eu P \limp \e\eu S)) \\
          3.&\a(\eu Q \limp   \a(\eu P \limp \e\eu S))
        \end{array}
        \) 
        \hfill
        \vspace{3mm}\par\noindent
        \(
        \begin{array}{rl}
          \multicolumn{2}{c}{$D. Between \(Q\) and \(R\)$} \\
          0.&\a((Q \land \e R) \limp 
                ((P \limp (\lnot R \U S)) \U R)) \\
          1.&\a((\eu Q \land \e\eu R \land \lnot\eu R) \limp
                   \o(((P \limp (\lnot\eu R \U S))\land\lnot\eu R)\U
                      (\eu R \land (P \limp Q)))) \\
          2.&\a((Q \land \e R) \limp
                ((\eu P \limp (\lnot R \U \eu S)) \U R)) \\
          3.&\a((\eu Q \land \e\eu R) \limp
                ((\eu P \limp (\lnot\eu R \U \eu S)) \U \eu R))
        \end{array}
        \)
        \hfill
        \vspace{3mm}\par\noindent
        \(
        \begin{array}{rl}
          \multicolumn{2}{c}{$E. After \(Q\) Until \(R\)$} \\
          0.&\a(Q \limp (P \limp (\lnot R \U S)) \W R) \\
          1.&\a(\eu Q \limp \o(((P \limp (\lnot\eu R \U S))\land\lnot\eu R) \W 
                               (\eu R \land (P \limp Q)))) \\
          2.&\a(\eu Q \limp (\eu P \limp (\lnot R \U \eu S)) \W R) \\
          3.&\a(\eu Q \limp (\eu P \limp (\lnot\eu R \U \eu S)) \W \eu R)
        \end{array}
        \)
        \(\begin{array}{l}\rule{0cm}{1mm}\end{array}\) 
      \end{minipage}
      }
      \caption{Formulations of the \textbf{Response} Pattern}
    \label{fig:patR}
  \end{center}
\end{figure*}

The four patterns that we did not extend are: \textbf{Bounded
Existence}, \textbf{Precedence Chain}, \textbf{Response Chain},
\textbf{Constrained Chain}. While we considered the \textbf{Bounded
Existence} pattern to be too convoluted to be useful in practice and
thus not worth the effort of extending, the other three patterns
were not extended for reasons that will become apparent in the
next Section.

\subsection{Discussion}

What is involved in adding events to a property?  Consider specifying
the absence pattern under the ``Between $Q$ and $R$ scope'' where $Q$
and $R$ are (up) edges.  The original formula is 
$$
\a( Q \land \e R \limp \lnot P \U R)
$$
This formula does not include $P$ when $Q$ and $R$ occur simultaneously.
This behavior is desired since the founding interval is
half open and thus becomes empty when the two ends
coincide. If we want to transform the condition and the interval bounds into
edges, we may be tempted to use the formula:
$$
\a(\eu Q \land \e\eu R \limp \lnot P \U \eu R)
$$
\begin{figure}[htb]
  \begin{center}
    \epsfig{file=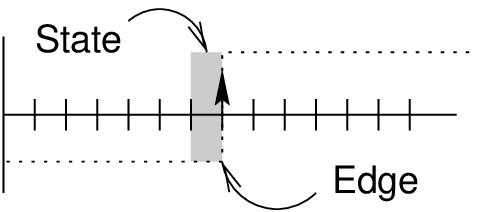} 
    \caption{Edge-detecting state}
    \label{fig:edge}
  \end{center}
\end{figure}
However, in order to effectively express properties containing edges, we
need to realize that an edge is detected just \emph{before} it
occurs, as illustrated in Figure~\ref{fig:edge}
That is, $\eu A$ becomes true
in the state where $A$ is false.

Thus, our formula has a problem:  we start testing $P$
\emph{before} the edge in $Q$ because this is when we detect $\eu Q$.
We need to fix this by testing $P$ \emph{after} the edge in $Q$:
$$
\a(\eu Q \land \e\eu R \limp\o (\lnot P \U \eu R))
$$
We fixed the above problem but introduced another: the new formula does not
work correctly when $\eu Q$ and $\eu P$ occur simultaneously.  This
happens because we make sure that $\eu R$ occurs one state too early.
We need to fix the antecedent by making sure that the interval
is non-empty:
$$
\a(\eu Q \land \lnot\eu R \land \e\eu R \limp\o (\lnot P \U \eu R))
$$
Unfortunately, the resulting formula is still incorrect: if $P$ and $\eu R$
occur simultaneously, then that occurrence of $P$ will be ignored since
$\eu R$ is detected in the state \emph{before} the edge. This is not
the intended behavior as the state before the edge is considered part
of the interval. We need to fix it one more time:
$$
\a(\eu Q \land \lnot\eu R \land \e\eu R \limp\o \lnot(\lnot\eu R \U P))
$$
or better yet:
$$
\a(\eu Q \land \lnot\eu R \land \e\eu R \limp\o(\eu R \P P))
$$

Note that we can avoid many complications of the sort discussed above
if we add the ``previous'' modality, $Y$. 
Using this new operator, we can
detect edges just \emph{after} they occur: $a \land \lnot Y a$ (see
Figure~\ref{fig:edge}).  Although having the ``previous'' operator can
potentially simplify a number of properties, it is currently not supported
by SPIN.

\section{Closure Under Stuttering}
\label{cus}

The extension of the pattern system presented in
Section~\ref{extensions}, is an important development, and we hope
that the resulting patterns provide real value to the end user.
Still, can practitioners use our extensions directly, without worrying
about any unexpected behavior?

The patterns that we have created in the previous section contain the
``next'' operator.  Thus, they may not be {\em closed under
stuttering}.  Intuitively, a formula is closed under stuttering when
its interpretation is not modified by transitions that leave the
system in the same state.  For example, a formula $\a a$ is closed
under stuttering because no matter how much we repeat states, we
cannot change the value of $a$.  On the other hand, the formula $\circ
a$ is not closed under stuttering.  We can see that by considering a
state sequence in which $a$ is true in the first state and false in
the second.  Then $\circ a$ is false if we evaluate it on this
sequence, and true if we repeat the first state.  

Closure under stuttering is an essential property of temporal formulas
to ensure basic separation between abstraction levels and to enable
powerful partial-order reduction algorithms utilized in mechanized
checking, e.g.~\cite{holzmann97}.  This property can be easily
guaranteed for a subset of LTL that does not include the ``next''
operator~\cite{lamport94}; however, events cannot be expressed in this
subset.  Determining whether an LTL formula is closed under stuttering
is hard: the problem has been shown to be
PSPACE-complete~\cite{peled96}.  A computationally feasible algorithm
which can identify a subclass of closed under stuttering formulas has
been proposed in~\cite{holzmann96} but have not yet been implemented;
without an implementation, one can not say how often the subclass of
formulas identified by the algorithm is encountered in
practice. Several temporal logics that try to solve the problem have
been proposed.  Such logics, e.g. TLA~\cite{lamport94} and
MTL~\cite{mokkedem94}, restrict the language so that all formulas
expressed in it are, by definition, closed under stuttering.  However,
it is not clear if these languages are expressive enough for practical
use.

In this section we briefly present a set of theorems that allow syntactic
reasoning about closure under stuttering in LTL formulas and show
how to apply them to our extensions of the pattern system.  For a more complete
treatment of closure under stuttering, please refer to~\cite{paun99b,paun99a}.

\subsection{Formal Definition}
The notation below is adopted from~\cite{hehner93}.
A sequence (or string) is a succession of
elements joined by semicolons.  For example, we write the sequence
comprised of the first five natural numbers, in order, as $0;1;2;3;4\ $
or, more compactly, as $0;..5$ (note the left-closed, right-open
interval).  We can obtain an item of the sequence by subscripting:
$(0;2;4;5)_2 = 4$.  When the subscript is a sequence, we obtain a
subsequence: $(0;3;1;5;8;2)_{1;2;3} = (0;3;1;5;8;2)_{1;..4} = 3;1;5$.
A state is modeled by a function that maps variables to their values,
so the value of variable $a$ in state $s_0$ is $s_0(a)$. We denote the
set of all infinite sequences of states as $\stseq$, and the set of
natural numbers as $\nat$. 

We say that an LTL formula $F$ is closed under stuttering when
its interpretation remains the same under state sequences that differ only
by repeated states. 
We denote an interpretation of formula $F$
in state sequence $s$ as $\interp{F}{s}$, and a closed under
stuttering formula as $\cus{F}$.  $\cus{F}$ is formally defined as follows:
\begin{definition} \label{eq:cus} $\cus{F} \Leq
  \A{s}{\stseq}\A{i}{\nat}\interp{F}{s} =
                                \interp{F}{(s_{0;..i};s_i;s_{i;..\infty})}$
\end{definition}
In other words, given any state sequence $s$, we can repeat any of its
states without changing the interpretation of $F$. Note that
$s_{0;..i};s_i;s_{i;..\infty}$ is a sequence of states that differs
from $s$ only by repeating a  state $s_i$.

\subsection{Properties}
Here we present several theorems that allow syntactic reasoning about
closure under stuttering.  First, we note that
\(\ed\) and \(\eu\) can be used
interchangeably when analyzing properties of the form
\[ \cus{A} \limp f(\eu A) \]
Thus, in what follows we will only discuss the $\eu$-edge.

We start with a few generic properties of closure under stuttering:
\begin{eqnarray}
(\nm{const}(a) \lor \nm{var}(a)) & \limp&  \cus{a} \label{cus-var}\\
\cus{A}                 &\leq&  \cus{\lnot A} \label{cus-not} \\
(\cus{A} \land \cus{B}) &\limp& \cus{A \land B} \label{cus-and}\\
(\cus{A} \land \cus{B}) &\limp& \cus{A \lor B} \label{cus-or}\\
(\cus{A} \land \cus{B}) &\limp& \cus{A \limp B} \label{cus-imp}\\
\cus{A}                 &\limp& \cus{\a A}     \\
\cus{A}                 &\limp& \cus{\e A}     \\
(\cus{A} \land \cus{B}) &\limp& \cus{A \U B}   
\end{eqnarray}
For example, (\ref{cus-var})-(\ref{cus-imp}) indicate that all
propositional formulas are closed under stuttering.  The above
properties do not include reasoning about formulas that contain the 
``next'' operator.  For those, we need the following theorem,
proven in~\cite{paun99b}:
\begin{theorem}[cus-main-thm]\label{thm:main}
\[\begin{array}{ll}
     & (\cus{A} \land \cus{B} \land \cus{C} \land 
       \cus{D} \land \cus{E} \land \cus{F}) \\
\limp& \\ 
     & \cus{(\lnot\eu A \lor \o B \lor C) \U 
            (\eu D \land \o E \land F)}
\end{array}\]
\end{theorem}
This theorem establishes an important relationship between the
``next'' operator, edges, and closure under stuttering. It gives rise
to a number of corollaries that we found to be useful in practice.

\begin{prop} \label{eq:cusEe}
\[
 (\cus{A} \land \cus{B} \land \cus{C})\limp\cus{\e(\eu A \land \o B \land C)} 
\]
\end{prop}
If we take $B=\ltop$ or $C=\ltop$ respectively, we obtain
two simplified versions:
\[
(\cus{A} \land \cus{B}) \limp \cus{\e(\eu A \land B)}
\]
and
\[
(\cus{A} \land \cus{B}) \limp \cus{\e(\eu A \land \o B)}
\]
These formulas
represent an \emph{existence} property: an event $\eu A$ must happen
and then $B$, evaluated before or after the event, should hold.

\begin{prop} \label{eq:cusAe}
\[
 (\cus{A} \land \cus{B} \land \cus{C}) \limp \cus{\a(\eu A \limp \o B \lor C)} 
\]
\end{prop}
is similar to Property~\ref{eq:cusEe}.
Its two simplified versions are
\[
(\cus{A} \land \cus{B}) \limp \cus{\a(\eu A \limp B)}
\]
and
\[
(\cus{A} \land \cus{B}) \limp \cus{\a(\eu A \limp \o B)}
\]
They express a \emph{universality} property: whenever an event $\eu
A$ happens, $B$, evaluated right before or right after the event, will
hold.


\subsection{Closure Under Stuttering and the Pattern System}
All properties of Figures~\ref{fig:patAb}-\ref{fig:patR} have been
shown to be closed under stuttering.  This was done using 
general rules of logic and properties identified above.
For example, consider checking a property
$$
\a(\eu Q \land \lnot\eu R \land \e\eu R \limp\o\lnot(\lnot\eu R \U P))
$$
for closure under stuttering.  The proof goes as follows:
\begin{widestproof}
     &\cus{\a(\eu Q\land\lnot\eu R \land\o\e\eu R \limp
              \o\lnot(\lnot\eu R \U P))} 
     \lhint{by rules of logic and LTL}
\Leq &\cus{\a(\eu Q\land\lnot\eu R \limp\o 
             (\lnot(\lnot\eu R\U P) \lor\lnot\e\eu R))}
     \lhint{by definition of $\eu$ and rules of logic}
\Leq &\cus{\a(\eu Q\land(R\lor\o\lnot R)\limp
             \o(\lnot(\lnot\eu R\U P)\lor\lnot\e\eu R))}
     \lhint{again, by rules of logic and LTL}
\Leq &\cus{\a((\eu Q\land R\limp\o(\lnot(\lnot\eu R\U P)\lor\lnot\e\eu R))) 
           \land \mbox{} \\
     &     \phantom{\cusleft}
           \a(\eu Q \limp \o(\lnot(\lnot\eu R\U P)\lor\lnot\e\eu R \lor R))}
     \lhint{distribute $\cus{{}}$ over the main conjunction}
\Lpmi&\cus{\a((\eu Q\land R \limp
                 \o(\lnot(\lnot\eu R\U P)\lor\lnot\e\eu R)))} \land \mbox{} \\
     &\cus{\a(\eu Q \limp\o(\lnot(\lnot\eu R\U P)\lor\lnot\e\eu R \lor R))}
     \lhint{we can use now Property~\ref{eq:cusAe} on both conjuncts}
\Lpmi&\cus{Q} \land \cus{R} \land
      \cus{\lnot(\lnot\eu R\U P)\lor\lnot\e\eu R} \land \mbox{} \\
     &\cus{Q} \land \cus{\lnot(\lnot\eu R\U P)\lor\lnot\e\eu R \lor R}
     \lhint{by rules of logic, (\ref{cus-not}) and (\ref{cus-and})}
\Lpmi&\cus{Q} \land \cus{R} \land \cus{\lnot\eu R\U P} \land 
      \cus{\e\eu R} \land \mbox{} \\
     &\cus{Q} \land \cus{\lnot\eu R\U P} \land \cus{\e\eu R} \land \cus{R}
     \lhint{by Theorem~\ref{thm:main} and Property~\ref{eq:cusEe} we get}
\Lpmi&\cus{Q} \land \cus{R} \land \cus{R} \land \cus{P} \land \cus{R}
\end{widestproof}
We have thus proved that
\[
(\cus{P}\land\cus{Q}\land\cus{R}) \limp 
      \cus{\a(\eu Q\land\lnot\eu R \land\e\eu R \limp
                                   \o\lnot(\lnot\eu R \U P))}\]
Although the property is fairly complicated, the proof is not long, is
completely syntactic, and each step in the proof is easy.  Such a
proof can potentially be performed by a theorem-prover with minimal
help from the user.

As we noted earlier, we did not extend all patterns to include events.
The reason is that  \textbf{Precedence Chain}, \textbf{Response Chain} and
\textbf{Constrained Chain} were not closed under stuttering even in
their state-based formulations.  Consider, for example, the 
\textbf{Response Chain}
pattern under the Global scope, formalized as
\begin{displaymath}
  \a((S \land \o\e T) \limp \o\e(T \land \e P))
\end{displaymath}
When we evaluate this formula on the state sequence $s$:
\begin{flushleft}
\(
\begin{array}{lcccl}
  & s_0 & s_1 & s_2 & \cdots \\
S &\top &\bot &\bot & \cdots \\
T &\top &\bot &\bot & \cdots \\
P &\bot &\bot &\bot & \cdots
\end{array}
\)
\end{flushleft}
we get $\top$ because the antecedent is always $\bot$. However, if we
stutter the first state $s_0$, we get the sequence $s_0;s$:
\begin{flushleft}
\(
\begin{array}{lccccl}
  & s_0 & s_0 & s_1 & s_2 & \cdots \\
S &\top &\top &\bot &\bot & \cdots \\
T &\top &\top &\bot &\bot & \cdots \\
P &\bot &\bot &\bot &\bot & \cdots
\end{array}
\)
\end{flushleft}
The interpretation of the formula on this sequence is now $\bot$
because the antecedent is $\top$ and the consequent is $\bot$ (since $\e
P$ is always $\bot$). As stuttering causes a change in the
interpretation of the formula, we can conclude that the formula is not
closed under stuttering.

\section{Conclusion}
\label{conclusion}
In this paper we developed a concept of edges and used it to extend
the pattern-based system of Dwyer et. al. to reasoning about events.
We have also presented a set of theorems that enable the syntax-based
analysis of a large class of formulas for closure under stuttering.
This class includes all LTL formulas of the patterns that appeared in
this paper.  Since research shows that patterns account for 90\% of
the temporal properties that have been specified so
far~\cite{dwyer99}, we believe that our approach is highly applicable
to practical problems.

The goal of the pattern-based approach is to enable practitioners to
easily codify complex properties into temporal logic. The extensions
presented in this paper allow them to express events easily and
effectively, without worrying about closure under stuttering.  We hope
that this work has moved us, as a community, one step closer to making
automatic verification more widely usable.

\newpage
\appendix

\section{Properties of Edges}
We list a few representative properties of edges here.  Their proofs
appear in~\cite{paun99b}.  \cite{paun99b} also contains 
a comprehensive study of the concept of edges.

Edges are related:
\begin{eqnarray}
  \eu\lnot A  &\leq& \ed A    \label{eq:euNot} \\
  \ed\lnot A  &\leq& \eu A    \label{eq:edNot} \\
  \ea\lnot A  &\leq& \ea A    \label{eq:eaNot}
\end{eqnarray}

\vspace{-0.1in}

\noindent Edges interact with the boolean operators as follows:
\begin{eqnarray}
\eu(A \land B) &\leq& (\eu A \land \o B)      \lor (\eu B \land \o A)\\
\eu(A \lor B)  &\leq& (\eu A \land \lnot B)   \lor (\eu B \land \lnot A)\\
\ed(A \land B) &\leq& (\ed A \land    B)      \lor (\ed B \land    A)\\
\ed(A \lor B)  &\leq& (\ed A \land \o\lnot B) \lor (\ed B \land \lnot\o A)
\end{eqnarray}

\vspace{-0.1in}

\noindent Edges interact with each other:
\begin{eqnarray}
  \ed\ed A &\leq&   \ed A\\
  \ed\eu A &\leq&   \eu A\\
  \eu\ed A &\leq& \o\ed A\\
  \eu\eu A &\leq& \o\eu A
\end{eqnarray}

\vspace{-0.1in}

\noindent Edges interact with temporal operators as follows:
\begin{eqnarray}
  \eu\o A     &\leq& \o\eu A \\
  \ed\o A     &\leq& \o\ed A \\
  \eu\a A     &\leq& \eu A \land \o\a A \\
  \ed\a A     &\leq& \lbot              \\
  \eu\e A     &\leq& \lbot              \\
  \ed\e A     &\leq& \ed A \land \o\a\lnot A \\
  \eu(A \U B) &\leq& \lnot(A \lor B) \land \o(A \U B) \\
  \ed(A \U B) &\leq& B \land \lnot\o(A \U B) 
\end{eqnarray}

\end{document}